\documentclass[twocolumn,floatfix, nofootinbib,prd,preprintnumbers,showpacs,showkeys,superscriptaddress,longbibliography]{revtex4-2}
\usepackage{slashed}
\usepackage{amsmath}
\usepackage{booktabs}
\usepackage{array}

\usepackage{graphicx} 
\usepackage{mathrsfs}
\usepackage{amssymb}
\usepackage{amsmath}
\date{\today}
\usepackage{mathtools}
\usepackage{comment}

\usepackage[mathscr]{euscript}
\usepackage{amsmath}
\usepackage{graphicx}
\usepackage{dcolumn}
\usepackage{bm}
\usepackage{epsfig}
\usepackage{amssymb,latexsym,mathrsfs}
\usepackage{graphicx}
\usepackage{color}
\usepackage{hyperref} 
\usepackage{float}
\usepackage{diagbox}

\usepackage{textgreek}

\usepackage{tikz}

\usepackage{amsmath}
\usepackage{physics}
\usepackage{csquotes} 
\newcommand*\diff{\mathop{}\!\mathrm{d}}

\hypersetup{
    colorlinks=true,
    linkcolor=red,
    citecolor=blue,
} 

\usepackage{amsmath}
\usepackage{amssymb}
\usepackage{subfigure}
\usepackage{url}
\usepackage{xcolor}
\usepackage{color}
\usepackage{soul}
\definecolor{amaranth}{rgb}{0.9, 0.17, 0.31}
\definecolor{purple(munsell)}{rgb}{0.62, 0.0, 0.77}
\definecolor{americanrose}{rgb}{1.0, 0.01, 0.24}
\definecolor{palatinateblue}{rgb}{0.15, 0.23, 0.89}
\definecolor{royalblue(web)}{rgb}{0.25, 0.41, 0.88}
\definecolor{hanpurple}{rgb}{0.32, 0.09, 0.98}
\definecolor{beaublue}{rgb}{0.74, 0.83, 0.9}
\definecolor{carminered}{rgb}{1.0, 0.0, 0.22}
\definecolor{brightpink}{rgb}{1.0, 0.0, 0.5}
\definecolor{vividviolet}{rgb}{0.62, 0.0, 1.0}
\hypersetup{ linktoc=all,
    colorlinks, linkcolor={palatinateblue},
    citecolor={brightpink}, urlcolor={amaranth}}

\usepackage{comment}

\newcommand{\be}{\begin{equation}}
\newcommand{\ee}{\end{equation}}
\newcommand{\bs}{\begin{split}} 
\newcommand{\bea}{\begin{eqnarray}}
\newcommand{\eea}{\end{eqnarray}}

\newcommand{\bes}{\begin{subequations}}
\newcommand{\ees}{\end{subequations}}

\newcommand{\bo}{\raise-1mm\hbox{\Large$\Box$}}

\usepackage{cleveref}
\crefname{equation}{Equation}{Equations}
\Crefname{equation}{Equation}{Equations}

\begin{document}

\title{Through the Looking-Glass, and What AdS Found There:\\ quantum particle production with a Whittaker spectrum} 
\author{Michael R. R. Good}
\email{michael.good@nu.edu.kz}
\affiliation{Physics Department \& Energetic Cosmos Laboratory, Nazarbayev University,\\
Astana 010000, Qazaqstan}
\affiliation{Leung Center for Cosmology and Particle Astrophysics,
National Taiwan University,\\ Taipei 10617, Taiwan}
\author{Eric V.\ Linder}
\email{evlinder@lbl.gov} 
\affiliation{Berkeley Center for Cosmological Physics \& Berkeley Lab, University of California,\\ Berkeley CA 94720, USA.}

\begin{abstract} 
Parity-inverted anti-de Sitter space -- ``flipped AdS'' -- is studied through the accelerating boundary correspondence of a moving mirror trajectory. The particle production exhibits positive energy flux and a finite total energy (both unlike AdS). The particle spectrum is of Whittaker form, with some similarities to a Planck thermal spectrum. We also derive the corresponding spacetime metric, with similarities to regular de Sitter space, but exhibiting a tower of repeated causal regions with horizons. 
\end{abstract} 

\date{\today} 

\maketitle

\section{Introduction} 

The thermal particle spectrum and constant energy flux emitted by the moving mirror (accelerating boundary) corresponding to de Sitter spacetime \cite{Good:2020byh} demonstrated that multiple different equations of motion can generate exact eternal Planck-distributed vacuum quanta, with the first such case identified by Carlitz and Willey \cite{carlitz1987reflections} as an analog to black hole evaporation \cite{Hawking:1974sw}. Leveraging the constant flux condition of the Möbius symmetry \cite{mobmir} inherent in the Davies-Fulling \cite{Davies:1976hi,Davies:1977yv} stress-tensor Schwarzian \cite{Birrell:1982ix,Fabbri}, a third solution emerged -- the anti-de Sitter (AdS) moving mirror -- distinguished by its eternally negative energy flux \cite{CAT_ads,Good:2017ddq}.

While the negative energy flux of the AdS trajectory is physically startling, it is nevertheless mathematically easy to obtain.  Even the AdS equation of motion itself may be obtained simply from the dS trajectory by introducing an imaginary acceleration parameter \cite{Good:2020byh},  $\kappa \to i\kappa$, reminiscent of a Wick rotation.

However, in 3+1 dimensions the particle production from an accelerating electron can only be fully captured by the moving mirror model \cite{Ievlev:2023inj,Lynch:2022rqy,Ievlev:2024zai,Ievlev:2023xzv} when a parity flip is used to account for both sides of the mirror. Only then does an accurate analysis of photon production and emitted power ensue.

Thus, investigating the radiation on the opposite side of the moving mirror has yielded valuable insights consistent with experimental observations. In particular when horizons are present, which are impossible for a massive electron, a careful analysis is required to understand particle creation and the energy carried, e.g.\ \cite{Good:2020uff}. Given the AdS mirror’s unique properties—negative energy flux, dual horizons, and asymptotic uniform acceleration—exploring its parity-flipped counterpart could provide further intriguing results. 
Moreover, since flipped AdS has no retarded 
time asymptote, we can test Walker's \cite{walker1985particle} consistency between the total stress-energy from the flux -- 
calculated from the mirror trajectory -- and the 
sum of quanta given by the beta Bogoliubov 
coefficients. 

In general, comparison between de Sitter, Anti-de Sitter, and flipped AdS behaviors could shed light on open questions regarding particle production, negative energy states, and asymptotic horizons or acceleration. 

In Section~\ref{sec:expan} we explore the dynamics, 
including the velocity, acceleration, and 
spacetime structure of flipped AdS. Section~\ref{sec:energy} calculates the emitted flux and total energy. 
We compute the beta Bogoliubov coefficients 
in Section~\ref{sec:spectrum}, producing the 
particle spectrum, and establish the consistency 
with the energy flux. Section~\ref{sec:vsplanck} contrasts the flipped 
AdS particle spectrum with the usual AdS 
spectrum, and Section~\ref{sec:spacetime} 
derives the spacetime metric equivalent to 
the flipped AdS mirror trajectory. We conclude 
in Section~\ref{sec:concl}.

\section{Flipped AdS Dynamics} \label{sec:expan} 

\begin{figure}[h]
    \centering
    \includegraphics[width=0.35\textwidth]{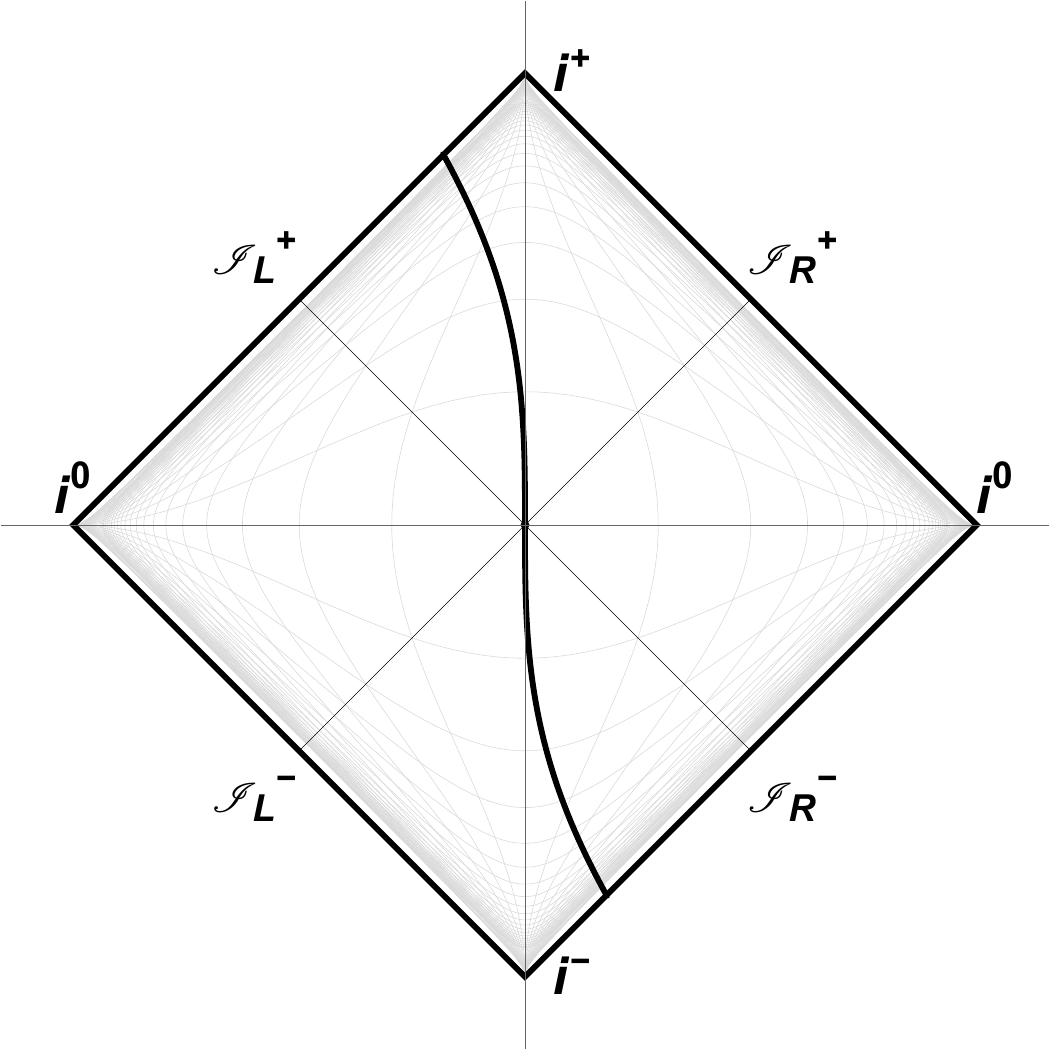}
    \caption{The trajectory Eq.~(\ref{traj}) is asymptotically light-like both in the past and future.  Notice the advanced time $v$ asymptotes, but retarded time $u \to \pm \infty$ is horizonless. 
    }
\label{fig:AdS_left_Penrose}
\end{figure}

\begin{figure}[h]
    \centering
    \includegraphics[width=0.35\textwidth]{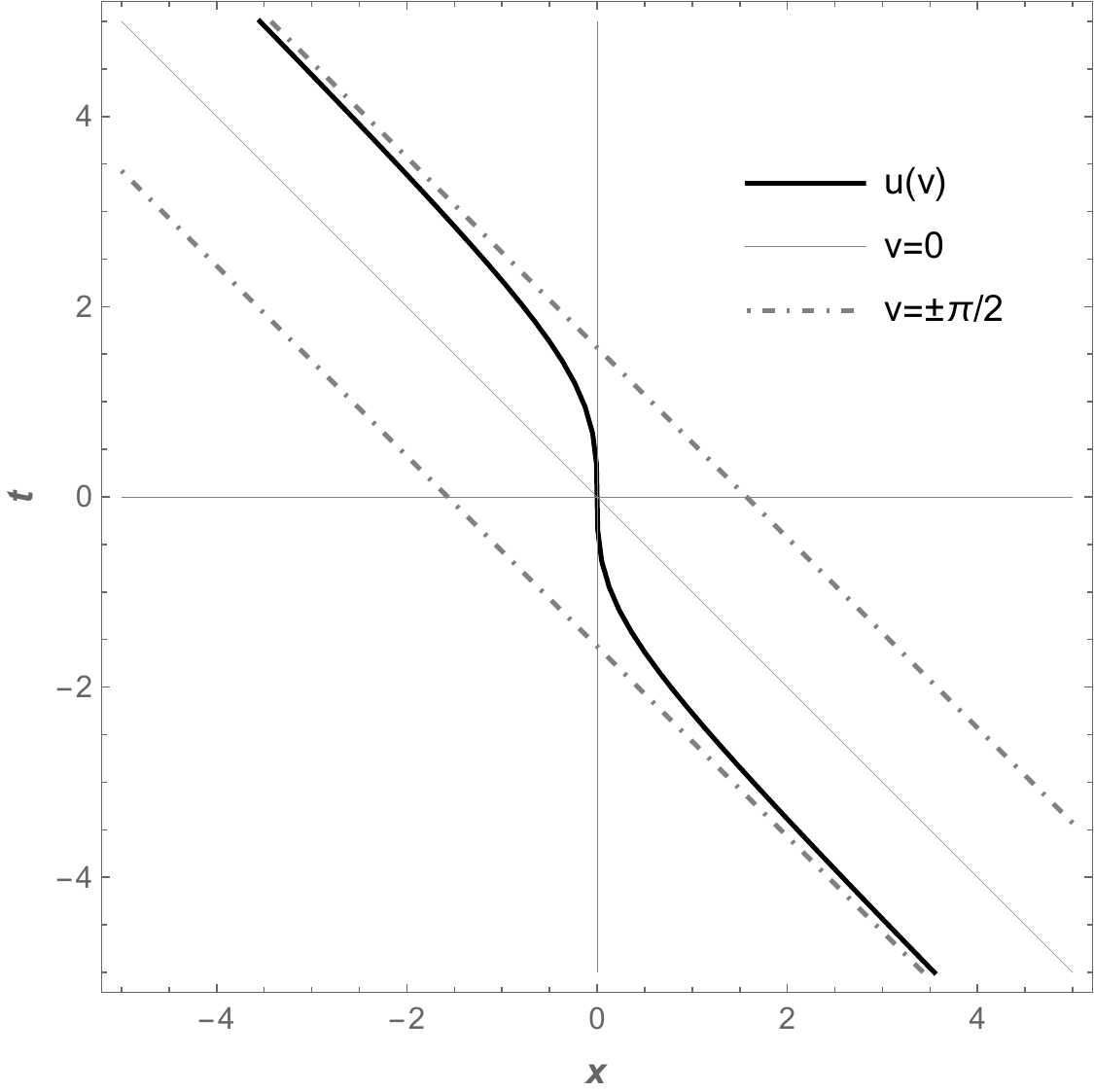}
    \caption{The trajectory Eq.~(\ref{traj}) in a spacetime plot demonstrates its motion is restricted between the advanced time $v \to \pm \pi/(2a)$ asymptotes (here we set $a=1$). 
    }
\label{fig:AdS_left_Spacetime}
\end{figure}

In light-cone coordinates, with 
retarded time $u = t-x$ and advanced time $v = t+x$, the standard AdS mirror trajectory is 
$u(v)=(1/a)\tan^{-1}(av)$. We define the 
flipped AdS mirror by a parity transformation, 
interchanging $u$ and $v$, so that 
\begin{equation}
  u(v) = \frac{1}{a} \tan (av) \quad \Leftrightarrow \quad v(u) = \frac{1}{a} \tan^{-1} (au)\ . \label{traj}
\end{equation}
Figure~\ref{fig:AdS_left_Penrose} shows the 
worldline in a conformal diagram, and 
Figure~\ref{fig:AdS_left_Spacetime} in a spacetime plot. Like the thermal mirror of de Sitter \cite{Good:2020byh} and the thermal anti-de Sitter mirror \cite{CAT_ads} (eternal negative energy flux), the flipped AdS mirror has some physically interesting traits all on its own. One aspect is that it is unbounded 
in the retarded time $u$ (going to $\pm\infty$), but bounded due to horizons 
in the advanced time $v$, with $v_H=\pm\pi/(2a)$.

The mirror velocity $\mathcal{V}\equiv\tanh\eta$ comes from the rapidity $\eta\equiv\frac{1}{2}\ln v'(u) \equiv -\frac{1}{2}\ln u'(v)$, giving 
\begin{equation}
\mathcal{V}=-1 + \frac{2}{a^2 u^2+2} = 1-\frac{4}{3+ \cos (2 a v)}\ , 
\end{equation} 
which in the $u\rightarrow \pm \infty$ limits or horizon limits of $v\rightarrow \pm \pi/(2a)$ give
\be \lim_{u\rightarrow\pm \infty}\mathcal{V}(u) = \lim_{v\rightarrow\pm \frac{\pi}{2a}}\mathcal{V}(v)= - 1\,.\ee
That is, the mirror (accelerating boundary) 
approaches the speed of light in the receding direction (to the left) away from the observer on the right-hand side near the horizon. 
At $x=t=0=u=v$, the velocity vanishes. 

The proper acceleration, $\alpha(u) = e^{-\eta} \eta'(u)$, or $\alpha(v) = e^\eta \eta'(v)$, is 
\begin{equation}
\alpha=-\frac{a^2 u}{\sqrt{a^2 u^2+1}} = -a \sin (a v)\,.  
\end{equation}
Now we can see the meaning of the quantity 
$a$; it is the asymptotically uniform 
(extremum) acceleration in the far past and far future:
\be 
\lim_{u\rightarrow\pm \infty}\alpha(u)=  \lim_{v\rightarrow\pm \frac{\pi}{2a}}\alpha(v) =\mp a\,.  \label{AdS_acceleration} 
\end{equation}

\section{Radiated Energy} \label{sec:energy} 

Accelerated boundaries, like black holes and spacetime horizons, give rise to quantum radiation of particles. 
The energy flux of radiated quanta from the accelerating boundary is \cite{Davies:1976hi, Birrell:1982ix, Fabbri} 
\begin{equation}
    F(u) = -\frac{1}{24\pi }\left(\frac{v'''(u)}{v'(u)}- \frac{3}{2} \left[\frac{v''(u)}{v'(u)]}\right]^2\right)\ , \label{eq:fluxu} 
\end{equation}
or in terms of $v$-coordinate is
\be
F(v) = \frac{1}{24 \pi} \frac{1}{[u'(v)]^2} \left( \frac{u'''(v)}{u'(v)} - \frac{3}{2} \left[\frac{u''(v)}{u'(v)} \right]^2 \right) .
\ee
These formulas reveal simple expressions for the energy flux from the flipped AdS mirror,  
\begin{equation}
    F = \frac{a^2}{12 \pi}\frac{1}{\left(a^2 u^2+1\right)^2} =  \frac{a^2}{12\pi} \cos ^4(a v)\,. \label{flux}
\end{equation} 

\begin{figure}[h]
    \centering
    \includegraphics[width=0.45\textwidth]{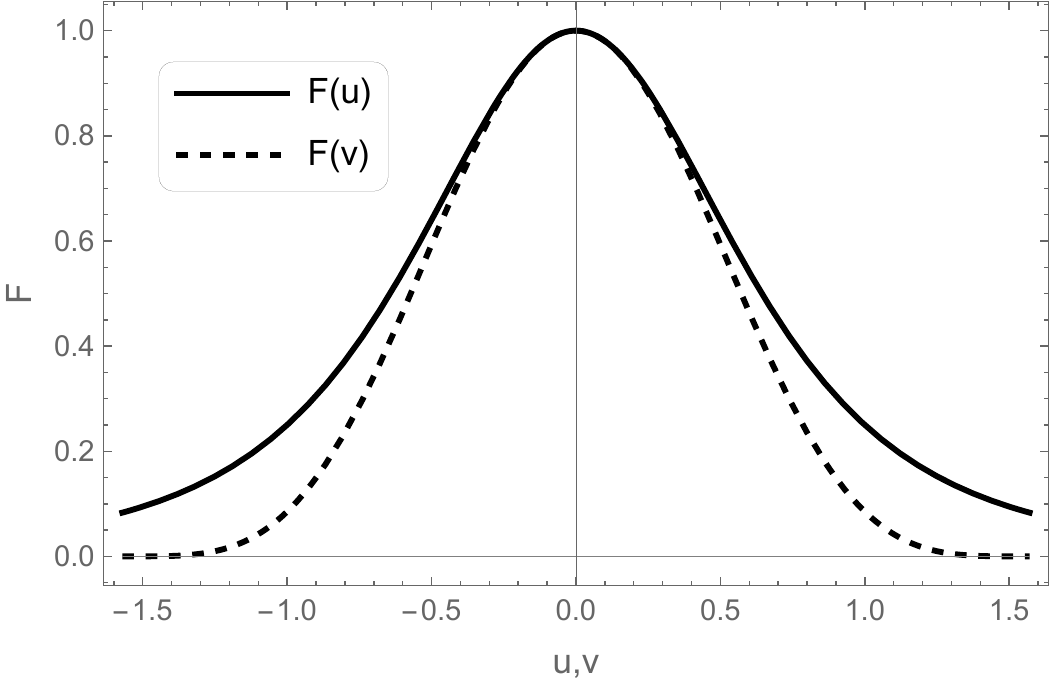}
    \caption{The energy flux, here normalized as $12\pi F/a^2$, is always positive, and the total energy emitted is finite.  
    }
\label{fig:AdS_F}
\end{figure}

The total energy emission is 
\begin{equation}
    E = \int_{-\infty}^{+\infty} F(u) \diff{u} = \int_{-\frac{\pi}{2a}}^{+\frac{\pi}{2a}} F(v) \frac{\diff{u}}{\diff{v}} \diff{v} = \frac{a}{24}\ . \label{total_energy}
\end{equation} 
The mirror accelerates forever, yet the total energy emission is finite. Note that at the 
asymptotes the flux dies off as $F\sim u^{-4}$ 
or $[|v|-\pi/(2a)]^4$. 
Figure~\ref{fig:AdS_F} plots the energy flux.

Having established the total energy emission from the flux in Eq.~(\ref{total_energy}), the next step is to derive the total energy from the particle creation at the quantum level, 
i.e.,\ through the creation/annihilation operators or beta Bogoliubov coefficients. 
For this, we recall that Walker \cite{walker1985particle} demonstrated complete consistency between particle and energy production predictions. The total energy released can be found by adding each quantum, 
\be E = \int_0^\infty \diff{p}\int_0^\infty\diff{q}\; p\, |\beta_{pq}|^2\ ,\label{sum}\ee 
where $p$ and $q$ are the outgoing and 
ingoing frequencies of the field modes subject to the accelerating boundary. 
The criteria for this 
to agree with the total stress-energy, 
\be E = \int_{-\infty}^\infty \diff{u} \; F(u)\,,\ee
require the absence of a retarded time asymptote in the mirror's trajectory (see the derivation of Eq.~5.4 in  \cite{walker1985particle}).  

The flipped AdS trajectory 
has no retarded time asymptote (cf.\ \cite{Good:2020byh,CAT_ads}). Thus, Walker's requirement is satisfied.  In this case, it is also encouraging that the radiated stress-energy, Eq.~(\ref{flux}),
\be F(u) = \frac{a^2}{12 \pi}\frac{1}{\left(a^2 u^2+1\right)^2}\ ,\label{fluxU}\ee
is always positive and finite, rather than the negative flux of standard AdS.  Adding quanta should intuitively work.  

Naively, it might be unsurprising that the total stress-energy released could be computed by adding each quantum. It is nevertheless essential to confirm this. Why? Demonstrating consistency would contribute valuable intuition because for standard (nonflipped) AdS the energy flux and total energy,
\be F = -\frac{\kappa^2}{48\pi},\qquad E = -\frac{\kappa}{24}, \label{NEF}\ee
are negative finite constants, which is interesting and physically unsettling.\footnote{For clarity, we use $\kappa$ for the AdS mirror and $a$ for the flipped AdS mirror; they would be related by $\kappa = 2a$.} Note that standard AdS does 
have a retarded time asymptote, and so fails the Walker criterion. (See Appendix~\ref{sec:apxasymp} for more detail.) 
The unique form of flipped AdS's positive and finite emission, 
Eq.~(\ref{fluxU}), is the radiative counterpart to the AdS mirror's time-independent emission of negative energy, Eq.~(\ref{NEF}).

\section{Whittaker Spectrum} \label{sec:spectrum} 

Now let us perform the energy summation over 
quantum modes and verify that it agrees with 
the stress-energy approach. 
We compute the beta Bogoliubov coefficient using the bounded integral in advanced time (see e.g.\ \cite{Birrell:1982ix}),  
\be \beta_{pq} = \frac{1}{2\pi} \sqrt{\frac{q}{p}}\,  \int_{-\frac{\pi}{2a}}^{+\frac{\pi}{2a}} \diff{v} \, e^{-iq v - ip u(v)}\ , 
\ee 
after an integration by parts, and the boundary term vanishing due to the asymptotic structure. 
Plugging in our trajectory Eq.~(\ref{traj})  gives
\be \beta_{pq} = \frac{1}{2\pi} \sqrt{\frac{q}{p}}\int_{-\frac{\pi }{2a}}^{+\frac{\pi }{2a}} \diff{v} \; \cos (\frac{p}{a} \tan (a v)+q v)\ ,\ee
where the $\sin$ integration is zero. 
We may write $V\equiv av$, $P=p/a$, $Q=q/a$, so that
\be \beta_{pq} = \frac{1}{a\pi} \sqrt{\frac{Q}{P}}\int_{0}^{+\frac{\pi }{2}} \diff{V} \; \cos \left(P \tan V+Q V\right). \label{beta_pq_integral}\ee 
Applying Gradshteyn \& Ryzhik 3.718.6 \cite{gradshteyn2014table},
\be
\int_{0}^{+\frac{\pi}{2}} \cos \left( P \tan x + Q x \right) \, \diff{x} = \frac{\pi}{2} 
\frac{W_{-\frac{Q}{2}, -\frac{1}{2}} (2P)}{ \Gamma \left( 1 - Q/2 \right)}\ ,
\ee
which gives a Whittaker function, except that when $Q$ is even then the integral and hence $\beta_{pq}=0$. 
Therefore, Eq.~(\ref{beta_pq_integral}) is equal to
\be \beta_{pq} = \frac{1}{2a}\sqrt{\frac{Q}{P}}\ \frac{W_{-\frac{Q}{2}, -\frac{1}{2}} (2P)}{ \Gamma \left( 1 - Q/2 \right)}\ . \ee
Note that $\beta_{pq}$ is real, a consequence 
of $u(v)$ being an odd function (see \cite{Good:2021dkh}), so 
\be |\beta_{pq}|^2 = \frac{1}{pq} \frac{\left[W_{-\frac{Q}{2},-\frac{1}{2}}(2 P)\right]^2}{ \left[\Gamma \left(-Q/2\right)\right]^2}\ .\ee 
Adding up the energy of each quantum as in Eq.~(\ref{sum}), 
\be E = \int_0^\infty \diff{p}\int_0^\infty\diff{q}\; p \; |\beta_{pq}|^2 = \frac{a}{24}\ , \ee
consistent with the energy emitted as derived by the stress-tensor, Eq.~(\ref{total_energy}). Quantum summing works, as it should in this case.

\section{Whittaker vs.\ Planck} \label{sec:vsplanck} 

The emitted particle spectrum follows from the beta Bogoliubov coefficients by 
\be 
n(p)=\int_0^\infty\diff{q}\; |\beta_{pq}|^2\ , 
\ee 
and the energy spectrum is simply $E(p)=p\,n(p)$. 

For the usual (``right hand side'' of the) AdS mirror, the 
particle spectrum is Planckian, i.e.\ thermal (though the energy flux is negative). Consider that the integral of  $E(p) = p\, n(p)$ for a Planck distribution with $T=a/(2\pi)$ gives the Planckian energy 
\be E = \int_0^{\infty }  p\, n(p) \, \diff{p} = \int_0^{\infty }  \frac{p/a}{e^{2\pi p/a}-1} \, \diff{p} = \frac{a}{24},\ee
which is the same as Eq.~(\ref{total_energy}). 

\begin{figure}[h]
   \centering
  \includegraphics[width=0.45\textwidth]{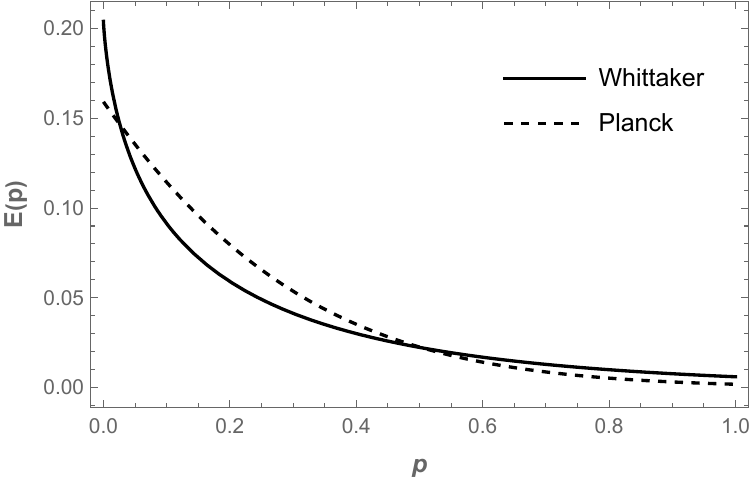}
    \caption{The energy of radiation  $E(p)$ as a function of frequency $p$ in units of $a$, for the Planck and Whittaker cases (cf., AdS, and flipped AdS). While the Whittaker case is not thermal, the total energy radiated is the same as the Planck energy with $T=a/(2\pi)$.}
\label{fig:Whittaker_vs_Planck}
\end{figure}

Despite the Planckian sum being the same as the Whittaker sum, the particle spectrum for the flipped AdS (or ``left hand side'') mirror is not Planckian but rather of  Whittaker form, 
\begin{align} 
E(p) &= \int_0^{+\infty} \frac{\mathrm{d}Q}{Q}\,\left[\frac{W_{-\frac{Q}{2},-\frac{1}{2}}(2 P)}{ \Gamma \left(-Q/2\right)}\right]^2\,
, & \text{Whittaker} \\
E(p) &= \frac{p/a}{e^{2\pi p/a} - 1}\ , & \text{Planck} 
\end{align}
Figure \ref{fig:Whittaker_vs_Planck}  compares these energy spectra.

Table~\ref{ta1} summarizes several related accelerating mirror trajectories and their properties, 
including the total flux, total energy, and whether quantum summing applies.

\begin{table*}[tbh!]
\caption{A comparison of de Sitter and Anti-de Sitter and related mirror models, usual and flipped.}
\centering
\setlength{\tabcolsep}{10pt} 
\begin{tabular}{|c|c|c|c|c|} 
\toprule
\textbf{Mirrors} & $v(u)$ & \textbf{Flux} & \textbf{Energy} & \textbf{Quantum Sum Eq.~(\ref{sum})} \\
\midrule
Carlitz-Willey & $\exp$ & $+\frac{\kappa^2}{48\pi}$ & $\infty$ & $\infty$ \\[1ex]
Carlitz-Willey Flipped & $\ln$ & $-\frac{\kappa^2}{48 \pi (\kappa u)^2}$ & $-\infty$ & \text{invalid} \\
\midrule
de Sitter & $\tanh$ & $+\frac{\kappa^2}{48\pi}$ & $\infty$ & $\infty$ \\[1ex]
de Sitter Flipped & $\tanh^{-1}$ & $-\frac{\kappa^2}{48 \pi} \left(\frac{\kappa^2 u^2}{4} + 1\right)^{-2}$ & $-\frac{\kappa}{48}$ & \text{invalid} \\
\midrule
Anti-de Sitter & $\tan$ & $-\frac{\kappa^2}{48\pi}$ & $-\frac{2\kappa}{48}$ & \text{invalid} \\[1ex]
Anti-de Sitter Flipped & $\tan^{-1}$ & $+\frac{\kappa^2}{48 \pi} \left(\frac{\kappa^2 u^2}{4} + 1\right)^{-2}$ & $+\frac{\kappa}{48}$ & \text{consistent!} \\
\bottomrule
\end{tabular}
\label{ta1}
\end{table*}

\section{Spacetime Metric} \label{sec:spacetime} 

Finally, we look at connecting the 
flipped AdS mirror to a spacetime, as 
can be done for the usual AdS and de 
Sitter cases. 
The accelerating boundary correspondence relates a mirror trajectory to a spacetime metric through matching conditions on null shells (see e.g.\ \cite{Good:2020byh} for explicit demonstration for the de Sitter case). 
Carrying this out for our flipped AdS case, 
we obtain the cosmological, i.e.\ spacetime, 
metric 
\begin{equation}
ds^2 = - f(r) dt^2 + \frac{dr^2}{f(r)} + r^2 d\Omega^2\,, 
\end{equation}
where $f(r) = \cos^2(r/L) = 1-\sin^2(r/L)$ 
and $L=\pi/(4a)$. 

This bears definite similarities to the de Sitter metric in static coordinates, which 
has $f_{\rm dS}(r)=1-(r/L)^2$ (see for example \cite{Balasubramanian:2001rb}). However, 
unlike the single cosmological horizon in de Sitter space at $r_{\rm dS}=L$, flipped AdS 
has an infinite sequence of horizons located at
\begin{equation}
r_n = \left(n + \frac{1}{2}\right) \pi L, \quad n \in \mathbb{Z}\ , 
\end{equation}
or with $\ell=2L/\pi=1/(2a)$, the locations of the horizons are at $\pm\ell$, $\pm3\ell$, $\pm5\ell$, etc.

Far from the horizon, however, $r\ll\pi/(4a)$, 
the spacetimes appear the same, $f(r)\approx 1-(r/L)^2$, with flipped AdS having 
an effective cosmological constant 
\begin{equation}
\Lambda = \frac{3}{L^2} = \frac{48 a^2}{\pi^2}\ .
\end{equation} 
At larger distances, the periodic nature of the flipped AdS $f(r)$ introduces a novel layered structure with repeating causal regions, suggestive of cyclic\footnote{Recall that 
the beta Bogoliubov coefficients vanish for $Q=q/a$ even, i.e.\ frequencies $q$ that are integer multiples of $\pi/(2L)$.} or emergent cosmological models, e.g., \cite{Steinhardt:2002ih,Vilenkin:1983xq,Ijjas:2021zwv}.

\section{Conclusions} \label{sec:concl} 

Anti-de Sitter space plays a central and intriguing 
role in understanding fundamental physics and spacetime. 
The accelerating boundary correspondence to AdS yields 
thermal but negative energy flux. Here we parity invert, 
or ``flip'', AdS and find interesting properties. The 
quantum energy emission is consistent, positive, and finite 
while the motion exhibits asymptotic uniform acceleration 
and no boundary asymptote in retarded time, though horizons 
in advanced time. Flipped AdS is the first explored particle 
production solution with these properties. 

Flipped AdS particle production has beta Bogoliubov coefficients 
that can be solved analytically, delivering a particle 
spectrum involving a Whittaker function. Moreover the summed 
quanta energy can be explicitly shown to agree with the total 
energy flux, as it should due to lack of boundary in retarded 
time. While the Whittaker spectrum is nonthermal, the integrated 
energy agrees with that of a thermal Planck spectrum. 
Since the usual AdS solution has negative energy flux, the 
appearance of the Whittaker spectrum in the flipped AdS case 
could yield insights into negative energy creation and 
understanding quantum field behavior under extreme conditions. 

Flipped AdS also has similarities to regular de Sitter space, 
with a cosmological, i.e.\ spacetime, metric that agrees with 
de Sitter far from the horizon. However flipped AdS actually 
has a tower of periodic spacetime horizons, suggestive of a 
cyclic universe. We can also preserve the flux properties of 
flipped AdS by a M\"obius transform (specifically inversion) 
but change the spacetime structure to obtain asymptotes in 
retarded time (see Appendix~\ref{sec:apxhidden}). These 
interconnections between AdS, flipped AdS, and de Sitter, 
and thermal, negative, and Whittaker particle creation, 
may be fruitful areas for further investigation.

\begin{acknowledgments}
M.G. is supported, in part, by the FY2024-SGP-1-STMM Faculty Development Competitive Research Grant (FDCRGP) no.201223FD8824 and SSH20224004 at Nazarbayev University in Qazaqstan. 
\end{acknowledgments}

\appendix 

\section{Retarded Time Asymptote and Standard vs Flipped AdS} \label{sec:apxasymp} 

To emphasize and clarify the role of the retarded time 
asymptote, or its lack, we consider 
an alternative approach to computing the emitted positive energy from flipped AdS. 
This method involves an integration by parts on the Davies-Fulling stress-energy, Eq.~(\ref{eq:fluxu}). The boundary term vanishes—as Walker predicts—because the mirror’s velocity asymptotically approaches $-1$, causing an advanced time asymptote, and the numerator tends to zero despite the proper acceleration remaining constant, leading to 
\be E = \frac{1}{48\pi} \int_{-\infty}^{+\infty} \left( \frac{v''(u)}{v'(u)} \right)^2 \, \diff{u}\ .\ee
Plugging in Eq.~(\ref{traj}) then yields
\be E = \frac{1}{48\pi} \int_{-\infty}^{+\infty}\frac{4 a^4 u^2}{\left(a^2 u^2+1\right)^2}\, \diff{u} = \frac{a}{24}\ .\ee

This contrasts with the non-vanishing boundary term from the standard (nonflipped) AdS mirror due to a velocity approach to +1, which causes a divergence from the denominator, ultimately causing negative energy.  The retarded time asymptote is the culprit and, overall, negative energy emission results. 
This negative sign spoils the ability to obtain the total emission by a sum of positive energies for each quantum from the standard AdS mirror.  No such problem exists for the flipped AdS mirror, as we showed in Section~\ref{sec:spectrum}.

\section{Hidden Horizons} \label{sec:apxhidden}

\begin{figure}[h]
   \centering
  \includegraphics[width=0.45\textwidth]{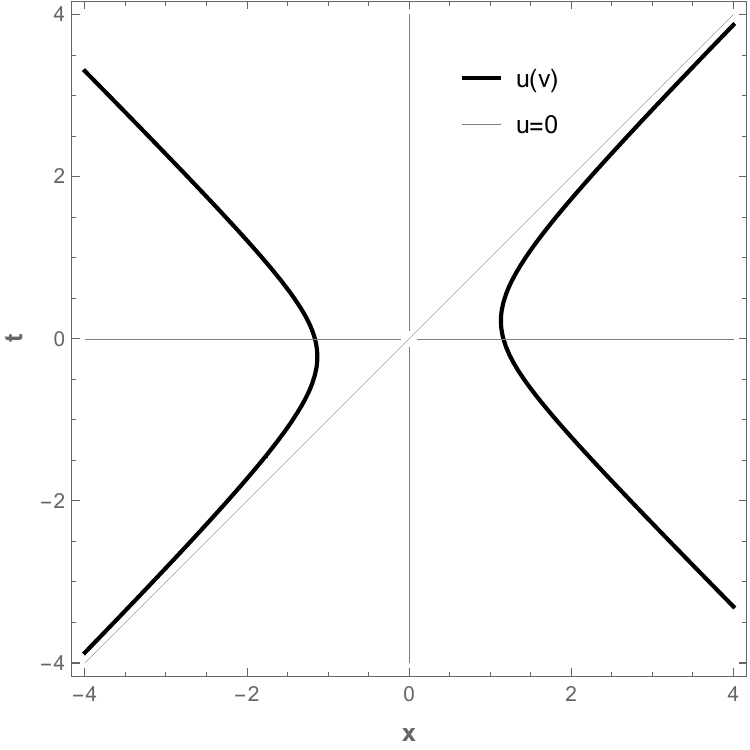}
    \caption{The inverted trajectory from a M\"obius transform of the flipped AdS trajectory, Eq.~(\ref{mobV}), is plotted in a spacetime diagram (for the principal value of $\tan^{-1}$).  The $u$-asymptote does not affect the energy flux and thus remains undetected by a measurement of the total energy. 
    }
\label{fig:mob}
\end{figure}

A M\"obius transform \cite{mobmir} of de Sitter's mirror \cite{Good:2020byh} gives the Carlitz-Willey \cite{carlitz1987reflections} trajectory. Therefore let us explore the M\"obius transform, $v(u) \rightarrow V(u)$, of the flipped AdS trajectory, specifically an inversion  \cite{mobmir}.  The inversion $V=-1/v$ gives the following trajectory with $v\to\pm \infty$ as $u \to 0^\mp$:
\be v(u) = \frac{1}{a} \tan^{-1} (au) \quad \longrightarrow \quad V(u) = \frac{-a}{\tan^{-1}(a u)}\ .\label{mobV}\ee 
Figure~\ref{fig:mob} shows the trajectory 
in a spacetime diagram. 

The flux emitted by an accelerating mirror following the $V(u)$ trajectory is the same as that emitted in the $v(u)$ case, Eq.~(\ref{flux}), consistent with M\"obius symmetry, so
\be E = \int_{-\infty}^{\infty} \diff{u}\; F(u) = \frac{a}{24}\ ,\ee
despite the $u$-asymptote. Nevertheless, due to the asymptote, the summation of quanta will not work.

\bibliography{main}

\begin{thebibliography}{22}%
\makeatletter
\providecommand \@ifxundefined [1]{%
 \@ifx{#1\undefined}
}%
\providecommand \@ifnum [1]{%
 \ifnum #1\expandafter \@firstoftwo
 \else \expandafter \@secondoftwo
 \fi
}%
\providecommand \@ifx [1]{%
 \ifx #1\expandafter \@firstoftwo
 \else \expandafter \@secondoftwo
 \fi
}%
\providecommand \natexlab [1]{#1}%
\providecommand \enquote  [1]{``#1''}%
\providecommand \bibnamefont  [1]{#1}%
\providecommand \bibfnamefont [1]{#1}%
\providecommand \citenamefont [1]{#1}%
\providecommand \href@noop [0]{\@secondoftwo}%
\providecommand \href [0]{\begingroup \@sanitize@url \@href}%
\providecommand \@href[1]{\@@startlink{#1}\@@href}%
\providecommand \@@href[1]{\endgroup#1\@@endlink}%
\providecommand \@sanitize@url [0]{\catcode `\\12\catcode `\$12\catcode `\&12\catcode `\#12\catcode `\^12\catcode `\_12\catcode `\%12\relax}%
\providecommand \@@startlink[1]{}%
\providecommand \@@endlink[0]{}%
\providecommand \url  [0]{\begingroup\@sanitize@url \@url }%
\providecommand \@url [1]{\endgroup\@href {#1}{\urlprefix }}%
\providecommand \urlprefix  [0]{URL }%
\providecommand \Eprint [0]{\href }%
\providecommand \doibase [0]{https://doi.org/}%
\providecommand \selectlanguage [0]{\@gobble}%
\providecommand \bibinfo  [0]{\@secondoftwo}%
\providecommand \bibfield  [0]{\@secondoftwo}%
\providecommand \translation [1]{[#1]}%
\providecommand \BibitemOpen [0]{}%
\providecommand \bibitemStop [0]{}%
\providecommand \bibitemNoStop [0]{.\EOS\space}%
\providecommand \EOS [0]{\spacefactor3000\relax}%
\providecommand \BibitemShut  [1]{\csname bibitem#1\endcsname}%
\let\auto@bib@innerbib\@empty
\bibitem [{\citenamefont {Good}\ \emph {et~al.}(2020)\citenamefont {Good}, \citenamefont {Zhakenuly},\ and\ \citenamefont {Linder}}]{Good:2020byh}%
  \BibitemOpen
  \bibfield  {author} {\bibinfo {author} {\bibfnamefont {M.~R.~R.}\ \bibnamefont {Good}}, \bibinfo {author} {\bibfnamefont {A.}~\bibnamefont {Zhakenuly}},\ and\ \bibinfo {author} {\bibfnamefont {E.~V.}\ \bibnamefont {Linder}},\ }\bibfield  {title} {\bibinfo {title} {{Mirror at the edge of the universe: Reflections on an accelerated boundary correspondence with de Sitter cosmology}},\ }\href {https://doi.org/10.1103/PhysRevD.102.045020} {\bibfield  {journal} {\bibinfo  {journal} {Phys. Rev. D}\ }\textbf {\bibinfo {volume} {102}},\ \bibinfo {pages} {045020} (\bibinfo {year} {2020})},\ \Eprint {https://arxiv.org/abs/2005.03850} {arXiv:2005.03850 [gr-qc]} \BibitemShut {NoStop}%
\bibitem [{\citenamefont {Carlitz}\ and\ \citenamefont {Willey}(1987)}]{carlitz1987reflections}%
  \BibitemOpen
  \bibfield  {author} {\bibinfo {author} {\bibfnamefont {R.~D.}\ \bibnamefont {Carlitz}}\ and\ \bibinfo {author} {\bibfnamefont {R.~S.}\ \bibnamefont {Willey}},\ }\bibfield  {title} {\bibinfo {title} {Reflections on moving mirrors},\ }\href {https://doi.org/10.1103/PhysRevD.36.2327} {\bibfield  {journal} {\bibinfo  {journal} {Phys. Rev. D}\ }\textbf {\bibinfo {volume} {36}},\ \bibinfo {pages} {2327} (\bibinfo {year} {1987})}\BibitemShut {NoStop}%
\bibitem [{\citenamefont {Hawking}(1975)}]{Hawking:1974sw}%
  \BibitemOpen
  \bibfield  {author} {\bibinfo {author} {\bibfnamefont {S.}~\bibnamefont {Hawking}},\ }\bibfield  {title} {\bibinfo {title} {{Particle Creation by Black Holes}},\ }\href {https://doi.org/10.1007/BF02345020} {\bibfield  {journal} {\bibinfo  {journal} {Commun. Math. Phys.}\ }\textbf {\bibinfo {volume} {43}},\ \bibinfo {pages} {199} (\bibinfo {year} {1975})}\BibitemShut {NoStop}%
\bibitem [{\citenamefont {Good}\ and\ \citenamefont {Linder}(2022)}]{mobmir}%
  \BibitemOpen
  \bibfield  {author} {\bibinfo {author} {\bibfnamefont {M.~R.~R.}\ \bibnamefont {Good}}\ and\ \bibinfo {author} {\bibfnamefont {E.~V.}\ \bibnamefont {Linder}},\ }\bibfield  {title} {\bibinfo {title} {{M\"obius mirrors}},\ }\href {https://doi.org/10.1088/1361-6382/ac60c3} {\bibfield  {journal} {\bibinfo  {journal} {Class. Quant. Grav.}\ }\textbf {\bibinfo {volume} {39}},\ \bibinfo {pages} {105003} (\bibinfo {year} {2022})},\ \Eprint {https://arxiv.org/abs/2108.07451} {arXiv:2108.07451 [gr-qc]} \BibitemShut {NoStop}%
\bibitem [{\citenamefont {Fulling}\ and\ \citenamefont {Davies}(1976)}]{Davies:1976hi}%
  \BibitemOpen
  \bibfield  {author} {\bibinfo {author} {\bibfnamefont {S.~A.}\ \bibnamefont {Fulling}}\ and\ \bibinfo {author} {\bibfnamefont {P.~C.~W.}\ \bibnamefont {Davies}},\ }\bibfield  {title} {\bibinfo {title} {Radiation from a moving mirror in two dimensional space-time: Conformal anomaly},\ }\href {https://royalsocietypublishing.org/doi/abs/10.1098/rspa.1976.0045} {\bibfield  {journal} {\bibinfo  {journal} {Proc. R. Soc. Lond. A}\ }\textbf {\bibinfo {volume} {348}},\ \bibinfo {pages} {393} (\bibinfo {year} {1976})}\BibitemShut {NoStop}%
\bibitem [{\citenamefont {Davies}\ and\ \citenamefont {Fulling}(1977)}]{Davies:1977yv}%
  \BibitemOpen
  \bibfield  {author} {\bibinfo {author} {\bibfnamefont {P.}~\bibnamefont {Davies}}\ and\ \bibinfo {author} {\bibfnamefont {S.}~\bibnamefont {Fulling}},\ }\bibfield  {title} {\bibinfo {title} {{Radiation from Moving Mirrors and from Black Holes}},\ }\href {https://doi.org/10.1098/rspa.1977.0130} {\bibfield  {journal} {\bibinfo  {journal} {Proc. R. Soc. Lond. A}\ }\textbf {\bibinfo {volume} {A356}},\ \bibinfo {pages} {237} (\bibinfo {year} {1977})}\BibitemShut {NoStop}%
\bibitem [{\citenamefont {Birrell}\ and\ \citenamefont {Davies}(1982)}]{Birrell:1982ix}%
  \BibitemOpen
  \bibfield  {author} {\bibinfo {author} {\bibfnamefont {N.~D.}\ \bibnamefont {Birrell}}\ and\ \bibinfo {author} {\bibfnamefont {P.~C.~W.}\ \bibnamefont {Davies}},\ }\href {https://doi.org/10.1017/CBO9780511622632} {\emph {\bibinfo {title} {{Quantum Fields in Curved Space}}}},\ Cambridge Monographs on Mathematical Physics\ (\bibinfo  {publisher} {Cambridge University Press},\ \bibinfo {address} {Cambridge, UK},\ \bibinfo {year} {1982})\BibitemShut {NoStop}%
\bibitem [{\citenamefont {Fabbri}\ and\ \citenamefont {Navarro-Salas}(2005)}]{Fabbri}%
  \BibitemOpen
  \bibfield  {author} {\bibinfo {author} {\bibfnamefont {A.}~\bibnamefont {Fabbri}}\ and\ \bibinfo {author} {\bibfnamefont {J.}~\bibnamefont {Navarro-Salas}},\ }\href {https://www.worldscientific.com/doi/abs/10.1142/p378} {\emph {\bibinfo {title} {Modeling Black Hole Evaporation}}}\ (\bibinfo  {publisher} {Imperial College Press},\ \bibinfo {year} {2005})\BibitemShut {NoStop}%
\bibitem [{\citenamefont {Myrzakul}\ and\ \citenamefont {Good}(2025)}]{CAT_ads}%
  \BibitemOpen
  \bibfield  {author} {\bibinfo {author} {\bibfnamefont {A.}~\bibnamefont {Myrzakul}}\ and\ \bibinfo {author} {\bibfnamefont {M.~R.~R.}\ \bibnamefont {Good}},\ }\bibfield  {title} {\bibinfo {title} {The cat and the mirror: classical acceleration temperature and the anti-de sitter dynamical casimir effect},\ }\href {https://doi.org/10.1142/S0217751X25430146} {\bibfield  {journal} {\bibinfo  {journal} {International Journal of Modern Physics A}\ }\textbf {\bibinfo {volume} {0}},\ \bibinfo {pages} {2543014} (\bibinfo {year} {2025})}\BibitemShut {NoStop}%
\bibitem [{\citenamefont {Good}\ and\ \citenamefont {Linder}(2018)}]{Good:2017ddq}%
  \BibitemOpen
  \bibfield  {author} {\bibinfo {author} {\bibfnamefont {M.~R.~R.}\ \bibnamefont {Good}}\ and\ \bibinfo {author} {\bibfnamefont {E.~V.}\ \bibnamefont {Linder}},\ }\bibfield  {title} {\bibinfo {title} {{Eternal and evanescent black holes and accelerating mirror analogs}},\ }\href {https://doi.org/10.1103/PhysRevD.97.065006} {\bibfield  {journal} {\bibinfo  {journal} {Phys. Rev. D}\ }\textbf {\bibinfo {volume} {97}},\ \bibinfo {pages} {065006} (\bibinfo {year} {2018})},\ \Eprint {https://arxiv.org/abs/1711.09922} {arXiv:1711.09922 [gr-qc]} \BibitemShut {NoStop}%
\bibitem [{\citenamefont {Ievlev}\ and\ \citenamefont {Good}(2024)}]{Ievlev:2023inj}%
  \BibitemOpen
  \bibfield  {author} {\bibinfo {author} {\bibfnamefont {E.}~\bibnamefont {Ievlev}}\ and\ \bibinfo {author} {\bibfnamefont {M.~R.~R.}\ \bibnamefont {Good}},\ }\bibfield  {title} {\bibinfo {title} {{Thermal Larmor Radiation}},\ }\href {https://doi.org/10.1093/ptep/ptae042} {\bibfield  {journal} {\bibinfo  {journal} {PTEP}\ }\textbf {\bibinfo {volume} {2024}},\ \bibinfo {pages} {043A01} (\bibinfo {year} {2024})},\ \Eprint {https://arxiv.org/abs/2303.03676} {arXiv:2303.03676 [gr-qc]} \BibitemShut {NoStop}%
\bibitem [{\citenamefont {Lynch}\ \emph {et~al.}(2024)\citenamefont {Lynch}, \citenamefont {Ievlev},\ and\ \citenamefont {Good}}]{Lynch:2022rqy}%
  \BibitemOpen
  \bibfield  {author} {\bibinfo {author} {\bibfnamefont {M.~H.}\ \bibnamefont {Lynch}}, \bibinfo {author} {\bibfnamefont {E.}~\bibnamefont {Ievlev}},\ and\ \bibinfo {author} {\bibfnamefont {M.~R.~R.}\ \bibnamefont {Good}},\ }\bibfield  {title} {\bibinfo {title} {{Accelerated electron thermometer: observation of 1D Planck radiation}},\ }\href {https://doi.org/10.1093/ptep/ptad157} {\bibfield  {journal} {\bibinfo  {journal} {PTEP}\ }\textbf {\bibinfo {volume} {2024}},\ \bibinfo {pages} {023D01} (\bibinfo {year} {2024})},\ \Eprint {https://arxiv.org/abs/2211.14774} {arXiv:2211.14774 [nucl-ex]} \BibitemShut {NoStop}%
\bibitem [{\citenamefont {Ievlev}\ \emph {et~al.}(2024{\natexlab{a}})\citenamefont {Ievlev}, \citenamefont {Good},\ and\ \citenamefont {Davies}}]{Ievlev:2024zai}%
  \BibitemOpen
  \bibfield  {author} {\bibinfo {author} {\bibfnamefont {E.}~\bibnamefont {Ievlev}}, \bibinfo {author} {\bibfnamefont {M.~R.~R.}\ \bibnamefont {Good}},\ and\ \bibinfo {author} {\bibfnamefont {P.~C.~W.}\ \bibnamefont {Davies}},\ }\bibfield  {title} {\bibinfo {title} {{Electron-mirror duality and thermality}},\ }\href {https://doi.org/10.1140/epjc/s10052-024-13557-0} {\bibfield  {journal} {\bibinfo  {journal} {Eur. Phys. J. C}\ }\textbf {\bibinfo {volume} {84}},\ \bibinfo {pages} {1159} (\bibinfo {year} {2024}{\natexlab{a}})},\ \Eprint {https://arxiv.org/abs/2405.06086} {arXiv:2405.06086 [quant-ph]} \BibitemShut {NoStop}%
\bibitem [{\citenamefont {Ievlev}\ \emph {et~al.}(2024{\natexlab{b}})\citenamefont {Ievlev}, \citenamefont {Good},\ and\ \citenamefont {Linder}}]{Ievlev:2023xzv}%
  \BibitemOpen
  \bibfield  {author} {\bibinfo {author} {\bibfnamefont {E.}~\bibnamefont {Ievlev}}, \bibinfo {author} {\bibfnamefont {M.~R.~R.}\ \bibnamefont {Good}},\ and\ \bibinfo {author} {\bibfnamefont {E.~V.}\ \bibnamefont {Linder}},\ }\bibfield  {title} {\bibinfo {title} {{IR-finite thermal acceleration radiation}},\ }\href {https://doi.org/10.1016/j.aop.2024.169593} {\bibfield  {journal} {\bibinfo  {journal} {Annals Phys.}\ }\textbf {\bibinfo {volume} {461}},\ \bibinfo {pages} {169593} (\bibinfo {year} {2024}{\natexlab{b}})},\ \Eprint {https://arxiv.org/abs/2304.04412} {arXiv:2304.04412 [gr-qc]} \BibitemShut {NoStop}%
\bibitem [{\citenamefont {Good}\ and\ \citenamefont {Abdikamalov}(2020)}]{Good:2020uff}%
  \BibitemOpen
  \bibfield  {author} {\bibinfo {author} {\bibfnamefont {M.}~\bibnamefont {Good}}\ and\ \bibinfo {author} {\bibfnamefont {E.}~\bibnamefont {Abdikamalov}},\ }\bibfield  {title} {\bibinfo {title} {{Radiation from an Inertial Mirror Horizon}},\ }\href {https://doi.org/10.3390/universe6090131} {\bibfield  {journal} {\bibinfo  {journal} {Universe}\ }\textbf {\bibinfo {volume} {6}},\ \bibinfo {pages} {131} (\bibinfo {year} {2020})},\ \Eprint {https://arxiv.org/abs/2008.08776} {arXiv:2008.08776 [gr-qc]} \BibitemShut {NoStop}%
\bibitem [{\citenamefont {Walker}(1985)}]{walker1985particle}%
  \BibitemOpen
  \bibfield  {author} {\bibinfo {author} {\bibfnamefont {W.~R.}\ \bibnamefont {Walker}},\ }\bibfield  {title} {\bibinfo {title} {Particle and energy creation by moving mirrors},\ }\href@noop {} {\bibfield  {journal} {\bibinfo  {journal} {Phys. Rev. D}\ }\textbf {\bibinfo {volume} {31}},\ \bibinfo {pages} {767} (\bibinfo {year} {1985})}\BibitemShut {NoStop}%
\bibitem [{\citenamefont {Gradshteyn}\ and\ \citenamefont {Ryzhik}(2014)}]{gradshteyn2014table}%
  \BibitemOpen
  \bibfield  {author} {\bibinfo {author} {\bibfnamefont {I.~S.}\ \bibnamefont {Gradshteyn}}\ and\ \bibinfo {author} {\bibfnamefont {I.~M.}\ \bibnamefont {Ryzhik}},\ }\href@noop {} {\emph {\bibinfo {title} {Table of Integrals, Series, and Products}}},\ \bibinfo {edition} {8th}\ ed.,\ edited by\ \bibinfo {editor} {\bibfnamefont {D.}~\bibnamefont {Zwillinger}}\ and\ \bibinfo {editor} {\bibfnamefont {V.~H.}\ \bibnamefont {Moll}}\ (\bibinfo  {publisher} {Academic Press},\ \bibinfo {address} {Cambridge, MA},\ \bibinfo {year} {2014})\BibitemShut {NoStop}%
\bibitem [{\citenamefont {Good}\ and\ \citenamefont {Linder}(2021)}]{Good:2021dkh}%
  \BibitemOpen
  \bibfield  {author} {\bibinfo {author} {\bibfnamefont {M.~R.~R.}\ \bibnamefont {Good}}\ and\ \bibinfo {author} {\bibfnamefont {E.~V.}\ \bibnamefont {Linder}},\ }\bibfield  {title} {\bibinfo {title} {{Light and Airy: a simple solution for relativistic quantum acceleration radiation}},\ }\href {https://doi.org/10.3390/universe7030060} {\bibfield  {journal} {\bibinfo  {journal} {Universe}\ }\textbf {\bibinfo {volume} {7}},\ \bibinfo {pages} {60} (\bibinfo {year} {2021})},\ \Eprint {https://arxiv.org/abs/2101.10576} {arXiv:2101.10576 [gr-qc]} \BibitemShut {NoStop}%
\bibitem [{\citenamefont {Balasubramanian}\ \emph {et~al.}(2001)\citenamefont {Balasubramanian}, \citenamefont {Horava},\ and\ \citenamefont {Minic}}]{Balasubramanian:2001rb}%
  \BibitemOpen
  \bibfield  {author} {\bibinfo {author} {\bibfnamefont {V.}~\bibnamefont {Balasubramanian}}, \bibinfo {author} {\bibfnamefont {P.}~\bibnamefont {Horava}},\ and\ \bibinfo {author} {\bibfnamefont {D.}~\bibnamefont {Minic}},\ }\bibfield  {title} {\bibinfo {title} {{Deconstructing de Sitter}},\ }\href {https://doi.org/10.1088/1126-6708/2001/05/043} {\bibfield  {journal} {\bibinfo  {journal} {JHEP}\ }\textbf {\bibinfo {volume} {05}},\ \bibinfo {pages} {043}},\ \Eprint {https://arxiv.org/abs/hep-th/0103171} {arXiv:hep-th/0103171} \BibitemShut {NoStop}%
\bibitem [{\citenamefont {Steinhardt}\ and\ \citenamefont {Turok}(2002)}]{Steinhardt:2002ih}%
  \BibitemOpen
  \bibfield  {author} {\bibinfo {author} {\bibfnamefont {P.~J.}\ \bibnamefont {Steinhardt}}\ and\ \bibinfo {author} {\bibfnamefont {N.}~\bibnamefont {Turok}},\ }\bibfield  {title} {\bibinfo {title} {{A Cyclic model of the universe}},\ }\href {https://doi.org/10.1126/science.1070462} {\bibfield  {journal} {\bibinfo  {journal} {Science}\ }\textbf {\bibinfo {volume} {296}},\ \bibinfo {pages} {1436} (\bibinfo {year} {2002})},\ \Eprint {https://arxiv.org/abs/hep-th/0111030} {arXiv:hep-th/0111030} \BibitemShut {NoStop}%
\bibitem [{\citenamefont {Vilenkin}(1983)}]{Vilenkin:1983xq}%
  \BibitemOpen
  \bibfield  {author} {\bibinfo {author} {\bibfnamefont {A.}~\bibnamefont {Vilenkin}},\ }\bibfield  {title} {\bibinfo {title} {{The Birth of Inflationary Universes}},\ }\href {https://doi.org/10.1103/PhysRevD.27.2848} {\bibfield  {journal} {\bibinfo  {journal} {Phys. Rev. D}\ }\textbf {\bibinfo {volume} {27}},\ \bibinfo {pages} {2848} (\bibinfo {year} {1983})}\BibitemShut {NoStop}%
\bibitem [{\citenamefont {Ijjas}\ and\ \citenamefont {Steinhardt}(2022)}]{Ijjas:2021zwv}%
  \BibitemOpen
  \bibfield  {author} {\bibinfo {author} {\bibfnamefont {A.}~\bibnamefont {Ijjas}}\ and\ \bibinfo {author} {\bibfnamefont {P.~J.}\ \bibnamefont {Steinhardt}},\ }\bibfield  {title} {\bibinfo {title} {{Entropy, black holes, and the new cyclic universe}},\ }\href {https://doi.org/10.1016/j.physletb.2021.136823} {\bibfield  {journal} {\bibinfo  {journal} {Phys. Lett. B}\ }\textbf {\bibinfo {volume} {824}},\ \bibinfo {pages} {136823} (\bibinfo {year} {2022})},\ \Eprint {https://arxiv.org/abs/2108.07101} {arXiv:2108.07101 [gr-qc]} \BibitemShut {NoStop}%
\end{thebibliography}%

\end{document}